\documentclass[final,twocolumn]{elsarticle}

\usepackage{lineno,hyperref}
\usepackage{braket}
\modulolinenumbers[5]

\journal{Journal of \LaTeX\ Templates}

\begin{document}

\begin{frontmatter}

\title{
Comment on ``Phase-space consideration on barrier transmission in a 
time-dependent variational approach with superposed wave packets (arXiv:2201.02966)'' 
}

\author[1]{N. Hasegawa}
\author[2]{K. Hagino}
\author[1,3]{Y. Tanimura}
\address[1]{Department of physics, Tohoku University, Sendai 980-8578, Japan}
\address[2]{Department of physics, Kyoto University, Kyoto 606-8502, Japan}
\address[3]{Graduate Program on Physics for the Universe, Tohoku University, Sendai 980-8578, Japan}

\begin{abstract}
We reply to the criticisms of our publication 
(N. Hasegawa, K. Hagino, and Y. Tanimura, Phys. Lett. B808, 135693 (2020)) 
made by A. Ono in his recent article, arXiv:2201.02966. 
\end{abstract}


\end{frontmatter}


In a recent publication, 
we employed the time-dependent generator coordinate method (TDGCM) 
and discussed a many-particle tunneling in the context of nuclear reaction\cite{HHT20}. 
In his preprint, Ono claims that 
our interpretation of the calculated results are unphysical and erroneous \cite{Ono}. 
The claim is based on the fact that 
the initial wave function employed in Ref. \cite{HHT20} 
has a broad momentum distribution. 

We do not fully agree with his point, at least with the form of 
wave functions considered in 
Ref. \cite{HHT20}, that is, 
Gaussian single-particle (s.p.) wave functions 
with a time-independent width. As is pointed out in Ref. \cite{Ono}, the relative 
wave function between two nuclei with such s.p. wave functions reads, 
\begin{equation}
\psi(x,t)=\sum_af_a(t)\,e^{-\nu_r\left(x-\frac{Z_a(t)}{\sqrt{\nu_r}}\right)^2},
\label{wf}
\end{equation}
where $f_a(t)$ is the weight factor for the Slater determinant $a$, $Z_a(t)$ is 
the center of the Gaussian wave packet, and $\nu_r$ is the time-independent 
width. 
We agree with Ref. \cite{Ono} that 
each Gaussian function in Eq. (\ref{wf}) has a broad momentum distribution if $\nu_r$ is taken to 
be 1.0 fm$^{-2}$ as in Ref. \cite{HHT20}. 
However, it is important to notice that the width $\nu_r$ is fixed in time in 
Eq. (\ref{wf}), and 
the wave function always has the same spatial width. 
This is in stark contrast to a quantum mechanical wave packet, whose spatial width 
broadens as a function of time according to its momentum distribution. 
Because of this, 
the Gaussian functions in Eq. (\ref{wf}) are 
compelled to behave classically. That is, even if the wave function (\ref{wf}) 
may have a broad momentum distribution, that will not be reflected fully in 
the time-dependent dynamics. This point was not properly mentioned in Ref. \cite{Ono}. 

Since each Gaussian function in Eq. (\ref{wf}) behaves classically, one may 
regard it as a 'classical (test) particle'. Under such mapping, the idea of the 
time-dependent 
generator coordinate method (\ref{wf}) is nothing more than that 
of the entangled trajectory molecular dynamics (ETMD) developed in quantum 
chemistry \cite{DM01,DZM03}. In ETMD, a quantum tunneling is simulated with 
many classical trajectories by 
making them entangled during the time evolution. This corresponds to taking 
a linear superposition of many Slater determinants in the time-dependent 
generator coordinate method. 
Such entanglement is a key to recover the quantum mechanical behavior of time-dependent 
dynamics. 

Of course, the energy still has a spreading in the time-dependent generator 
coordinate method, since each Gaussian wave function in Eq. (\ref{wf}) has 
a different mean energy. In our opinion, the energy spreading inherent in each Gaussian 
wave function should not fully be included 
in estimating the total energy spreading, 
as it does not contribute to the dynamics when a single wave packet is involved. 
That is, one should make a clear distinction between the energy spreading inherent in 
single-particle wave functions and that due to a distribution of initial 
conditions. 
As the former does not contribute to the dynamics when a single wave packet is 
considered, 
it is not trivial to estimate 
appropriately the energy spreading of initial wave packet for the TDGCM when 
a Gaussian function with a fixed width is employed. 
In our particular example shown in Ref. \cite{HHT20}, the mean energy and the 
energy variance due to the 
superposition of two Gaussian functions are 
0.11 MeV and 6.5$\times 10^{-3}$ MeV, respectively, 
which are compared to the average barrier height of 0.13 MeV. 
Notice that the energy variance is much smaller than the barrier height. 

Another claim of Ono is that the energy of the transmitted wave packet is  
considerably 
different from that of the initial wave packet. It might be tempting to 
regard the energy of each Slater determinant in the final state as the final 
energy of each reaction 
process. However, a care must be taken in this interpretation, due to 
the well known spurious cross-channel correlation in the time-dependent 
Hartree-Fock theory \cite{RS80,AK81}. This problem may not be cured 
completely if the number 
of Slater determinants superposed in the wave function is small. 
As in ETMD, only the total energy has a clear physical meaning 
in considering quantum tunneling, 
rather than the energy of each ``test particle''. 

\section*{Acknowledgements}
This work was supported in part by the Graduate Program on Physics for 
the Universe at Tohoku university, and in part by 
JSPS KAKENHI Grant Number JP19K03861.

\end{document}